\documentclass[10pt, conference, compsocconf]{IEEEtran}
\usepackage{bookmark}
\usepackage{graphicx}
\usepackage{amssymb}
\usepackage{color}
\usepackage{syntax}
\usepackage{listings}
\let\labelindent\relax
\usepackage{enumitem, pdfcomment}
\usepackage{mathtools}
\usepackage{algorithm}
\usepackage{comment}
\usepackage{subfig}
\usepackage{color}
\usepackage{listings}
\usepackage{stfloats}
\usepackage{float}
\newenvironment{ppl}{\small\ttfamily}{}
\ifCLASSINFOpdf
\else
\fi
\begin{document}

\title{Workflow Partitioning and Deployment on the Cloud using Orchestra}
\author{\IEEEauthorblockN{Ward Jaradat, Alan Dearle, and Adam Barker}
\IEEEauthorblockA{School of Computer Science, University of St Andrews, 
 North Haugh, St Andrews, Fife, KY16 9SX, United Kingdom\\
\{ward.jaradat, alan.dearle, adam.barker\}@st-andrews.ac.uk}
}
\maketitle
\begin{abstract}
Orchestrating service-oriented workflows is typically based on a design model that routes both data and control through a single point -- the centralised workflow engine.
This causes scalability problems that include the unnecessary consumption of the network bandwidth, high latency in transmitting data between the services, and performance bottlenecks.
These problems are highly prominent when orchestrating workflows that are composed from services dispersed across distant geographical locations.
This paper presents a novel workflow partitioning approach, which attempts to improve the scalability of orchestrating large-scale workflows.
It permits the workflow computation to be moved towards the services providing the data in order to garner optimal performance results.
This is achieved by decomposing the workflow into smaller sub workflows for parallel execution, and determining the most appropriate network locations to which these sub workflows are transmitted and subsequently executed.
This paper demonstrates the efficiency of our approach using a set of experimental workflows that are orchestrated over Amazon EC2 and across several geographic network regions.
\end{abstract}

\IEEEpeerreviewmaketitle

\section{Introduction}\label{Introduction}
Service workflows represent the automation of services during which data is passed between the services for processing.
Typically, workflows are orchestrated based on a centralised design model that provides control over the workflow, supports process automation, and encapsulates the workflow logic in a central location at which its execution takes place.
There are several languages used to describe service workflows such as the Business Process Execution Language (BPEL) \cite{BPEL}, which has been accepted as a standard service orchestration language.
The Simple Conceptual Unified Flow Language (SCUFL) is an example of a language for specifying large-scale workflows such as those seen in scientific applications \cite{CharSci}.
It is supported by the Taverna workbench and is typically executed using a workflow engine known as Freefluo \cite{Taverna}.
However, workflow management systems of this kind route both data and control through a single point, which causes scaling problems including the unnecessary consumption of network bandwidth, high latency in transmitting data between the services, and performance bottlenecks.\\

Scientific workflows can be composed from services that may be dispersed across distant geographical locations.
Determining the most appropriate location at which to execute the workflow logic becomes difficult as the number of geographically distributed services increases.
Most workflow management approaches rely on data placement strategies that attempt to move the data closer to locations at which the computation takes place \cite{DP1, DP2, DP3, DP4, DP5}.
This involves a set of activities related to data transfer, staging, replication, management and allocation of resources.
However, the distribution of large portions of data between the services and across long distances through the centralised engine can affect the data transfer rate, increase the execution time, risk overwhelming the storage resources at execution sites, and degrade the overall workflow performance.
Recent research efforts show interest in using Infrastructure as a Service (IaaS) clouds that provide on-demand computational services to support cost-efficient workflow management \cite{C1, C2}, but do not examine how the geographical location of services can affect the workflow performance.\\

The principal contribution of this paper is a partitioning approach that permits a workflow to be decomposed into smaller sub workflows for parallel execution on the cloud.
It determines the most appropriate locations to which these sub workflows are transmitted and subsequently executed.
Unlike existing approaches that depend on data placement, our approach permits the workflow computation to be moved closer to the services providing the data.
Through adopting this approach, distributed workflow engines can collaborate together to execute the overall workflow.
Each engine is carefully selected to execute a smaller part of the workflow within short network distance to the services.
For instance, an engine may be selected if it is hosted in the same network region where the services are resident in a cloud-based environment.
Our approach relies on collecting Quality of Service (QoS) information that represents the network delay (e.g. network latency and bandwidth) with a combination of heuristics to select the nearest engines to the services.
It is hypothesised that this improves the workflow performance by reducing the overall data transfer among the services.\\

Previously we created a distributed architecture for executing service workflows  \cite{Architecture}, 
which relies on a high-level language \cite{Language} for specifying workflows.
However, the published articles relating to these works do not discuss our workflow partitioning approach.
This paper presents a refined design of our architecture, evaluates our approach accordingly, and compares it to existing works.
In order to investigate the benefits of our approach, we use a set of experimental workflows that are executed over Amazon EC2 and across several geographic network regions.
These workflows are based on dataflow patterns that are commonly used to compose large-scale scientific workflows \cite{barker2007scientific}, which include the pipeline, distribution and aggregation patterns.\\

The rest of this paper is organised as follows:
Section \ref{workflowexamplesection} presents a simple workflow example that is used throughout the paper to explain our approach.
Section \ref{WorkflowPartitioning} presents our workflow partitioning approach.
Section \ref{WorkflowPartitioningExample} discusses a workflow partitioning example.
Section \ref{Evaluation} discusses and evaluates our approach implementation.
Section \ref{RelatedWorks} reviews related works.
Finally, section \ref{Conclusion} summarises our work achievements and states future research directions.

\section{Workflow Example}\label{workflowexamplesection}

\begin{figure}[h]
\centerline{\includegraphics{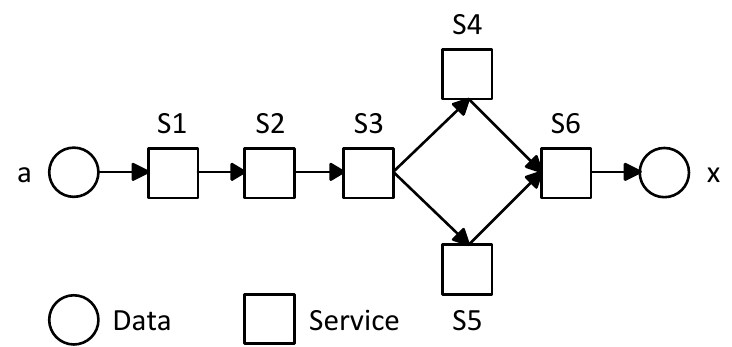}}
\caption{A Directed Acyclic Graph (DAG) workflow.}
\label{workflowfigure}
\end{figure}

Our approach relies on a new high-level functional data coordination language for the specification of workflows known as Orchestra.
It separates the workflow logic from its execution, and permits a workflow architect (e.g. scientist, engineer) to design a workflow without knowledge of how it is executed.
Orchestra allows a workflow to be composed as a Directed Acyclic Graph (DAG) that supports common dataflow patterns, and provides succinct abstractions for defining the services and coordinating the dataflow between them.
This section provides a simple workflow example that is used throughout this paper to explain our approach.
Figure \ref{workflowfigure} shows its structure, where the input \begin{ppl}a\end{ppl} is used to invoke service \begin{ppl}S1\end{ppl}, which produces an output that is used to invoke \begin{ppl}S2\end{ppl} whose output is then passed to \begin{ppl}S3\end{ppl}.
The output of \begin{ppl}S3\end{ppl} is used to invoke both \begin{ppl}S4\end{ppl} and \begin{ppl}S5\end{ppl}, whose outputs are used as inputs for \begin{ppl}S6\end{ppl}, which produces the final workflow output \begin{ppl}x\end{ppl}.

Listing \ref{listing1} presents the specification of this workflow using our language where the workflow name \begin{ppl}example\end{ppl} is defined in line 1 using the \begin{ppl}\textbf{workflow}\end{ppl} keyword.
The \begin{ppl}\textbf{description}\end{ppl} keyword is used to declare identifiers for a service description documents, each of which can be located using a URL through lines 2-7.
This permits the compiler to retrieve information about the services, their operations and associated types for syntax analysis.
The \begin{ppl}\textbf{service}\end{ppl} keyword is used to declare the service identifiers \begin{ppl}s1\end{ppl}, \begin{ppl}s2\end{ppl}, \begin{ppl}s3\end{ppl}, \begin{ppl}s4\end{ppl}, \begin{ppl}s5\end{ppl} and \begin{ppl}s6\end{ppl} through lines 8-13.
Similarly, the service ports \begin{ppl}p1\end{ppl}, \begin{ppl}p2\end{ppl}, \begin{ppl}p3\end{ppl}, \begin{ppl}p4\end{ppl}, \begin{ppl}p5\end{ppl} and \begin{ppl}p6\end{ppl} are declared using the \begin{ppl}\textbf{port}\end{ppl} keyword through lines 14-19.
The \begin{ppl}\textbf{input}\end{ppl} and \begin{ppl}\textbf{output}\end{ppl} keywords define the workflow interface, which provides an input \begin{ppl}a\end{ppl} and an output \begin{ppl}x\end{ppl} of the same type through lines 20-23.\\

\begin{lstlisting}[keywords={workflow, specification, engine, description, service, port, engine, is, input, output, int, any}, caption={Specification of the workflow in figure \ref{workflowfigure}.}, label={listing1}, frame=single]
01 workflow example
02 description d1 is http://ward.host.cs.st-andrews.ac.uk/documents/service1.wsdl
..
07 description d6 is http://ward.host.cs.st-andrews.ac.uk/documents/service6.wsdl
08 service s1 is d1.Service1
..
13 service s6 is d6.Service6
14 port p1 is s1.Port1
..
19 port p6 is s6.Port6
20 input: 
21    int a
22 output:
23    int x
24 a -> p1.Op1
25 p1.Op1 -> p2.Op2
26 p2.Op2 -> p3.Op3
27 p3.Op3 -> p4.Op4, p5.Op5
28 p4.Op4 -> p6.Op6.par1
29 p5.Op5 -> p6.Op6.par2
30 p6.Op6 -> x
\end{lstlisting}

\begin{figure*}[b]
\centerline{\includegraphics{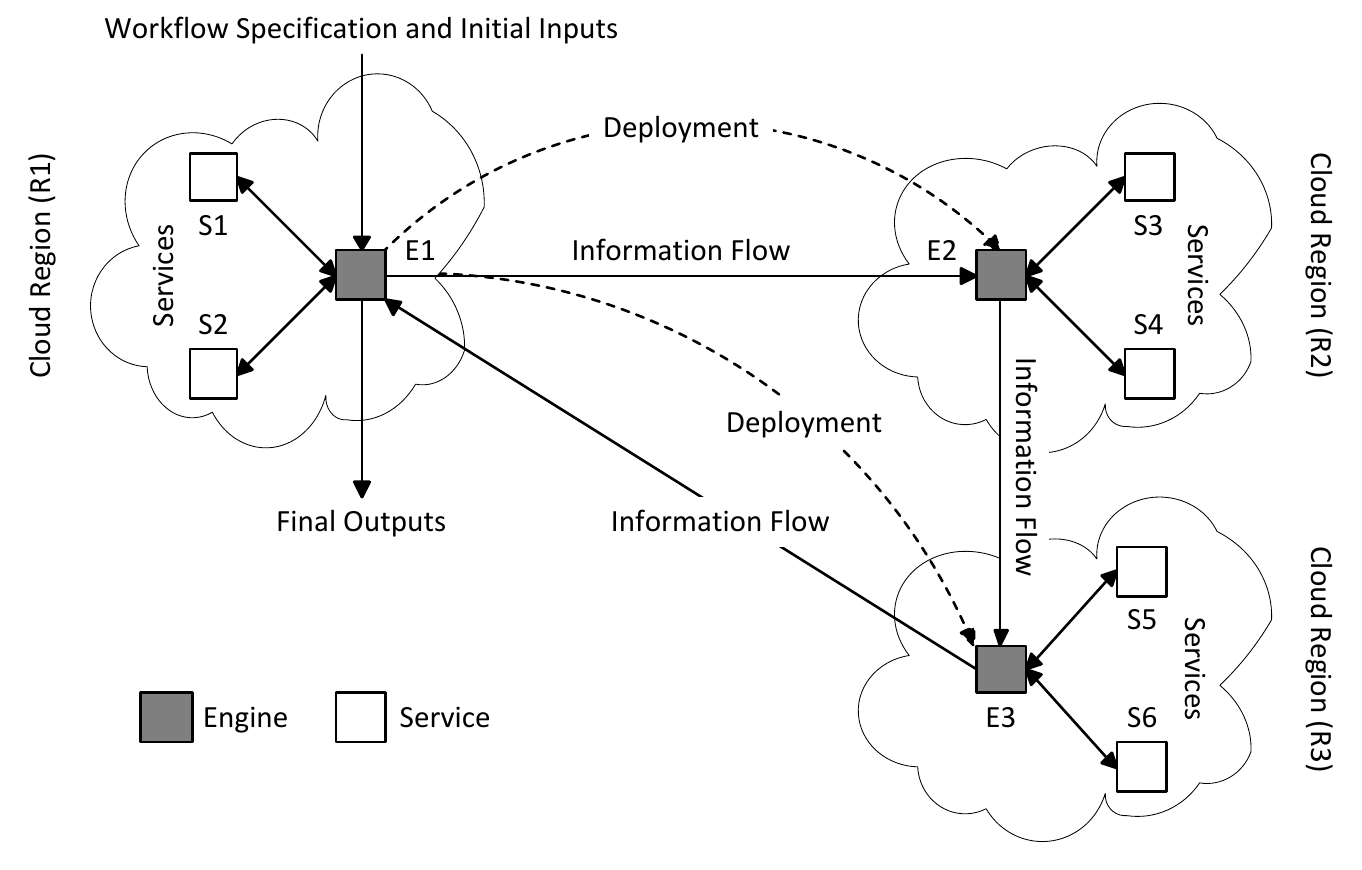}}
\caption{Overview of our distributed service orchestration architecture, and the interactions between distributed workflow engines. This diagram shows an arbitrary placement of services within cloud regions based on the example in figure \ref{workflowfigure}.}
\label{architecture}
\end{figure*}

Our language supports common dataflow patterns by specifying service invocations and the data passed to them.
Each service invocation consists of a port identifier and an associated operation separated by a dot symbol.
The output of a particular service invocation can be associated with an identifier, or passed directly to another service invocation to create a service composition.
The arrow symbol indicates the direction of the data to or retrieved from service invocations.
The following dataflow patterns are specified in listing \ref{listing1}:

\begin{itemize}
\item \textbf{Pipeline pattern:}
This pattern is used for chaining several services together, where the output of a particular service is used as an input to invoke another.
For instance, \begin{ppl}a\end{ppl} is used to invoke \begin{ppl}p1.Op1\end{ppl} whose result is passed directly to \begin{ppl}p2.Op2\end{ppl}, which in turn produces a result that is passed to \begin{ppl}p3.Op3\end{ppl} through lines 24-26.
\item \textbf{Data distribution pattern:}
This pattern is used to transmit several identical copies of a particular service output to multiple services.
For instance, the invocation result of \begin{ppl}p3.Op3\end{ppl} is used to invoke both \begin{ppl}p4.Op4\end{ppl} and \begin{ppl}p5.Op5\end{ppl} in line 27.
This finite sequence of invocations is the simplest parallel data structure in our language where each invocation is executed concurrently.
\item \textbf{Data aggregation pattern:}
The results of several service invocations may be passed as individual input parameters to one particular service using this pattern.
For instance, the results of both \begin{ppl}p4.Op4\end{ppl} and \begin{ppl}p5.Op5\end{ppl} are used as input parameters \begin{ppl}par1\end{ppl} and \begin{ppl}par2\end{ppl} respectively to invoke operation \begin{ppl}p6.Op6\end{ppl} through lines 28-29.
Finally, \begin{ppl}x\end{ppl} represents the result of \begin{ppl}p6.Op6\end{ppl} in line 30.
\end{itemize}

\section{Overview of Approach}\label{WorkflowPartitioning}

In order to realise our approach we created a fully distributed orchestration architecture.
Unlike existing orchestration technology where the locus of control is represented by a centralised engine that holds the decision logic of the workflow, the notion of a single locus of control does not exist in our architecture.
During the workflow execution, the decision logic can be found at one or several engines.
Figure \ref{architecture} shows an overview of our architecture that consists of distributed workflow engines.
These engines collaborate together to complete the workflow execution.
For instance, a workflow engine may be responsible for analysing and partitioning a workflow specification into smaller sub workflows.
These sub workflows may be deployed onto remote engines for execution.
Each engine then exploits connectivity to a group of services by invoking them or composing them together, retrieving invocation results, and forwarding relevant information to remote engines as necessary.
The following sections discuss the compilation and partitioning of a workflow specification, deployment of sub workflows and monitoring their execution.

\subsection{Compilation}

Our approach uses a recursive descent compiler that analyses a workflow specification to ensure its correctness.
It does not generate machine code representation from the workflow specification, but constructs an executable graph-based data structure instead, which consists of vertices that represent service invocations with edges between them as data dependencies.
The components of this data structure can be distributed to remote workflow engines at arbitrary network locations.
This permits a workflow to be maintained upon its distribution such that it can be refactored for optimisation purposes during run-time.

\subsection{Partitioning}\label{PartitioningSection}

Our workflow partitioning approach consists of several phases that include workflow decomposition, placement analysis, and composition of sub workflows.\\

\subsubsection{Decomposition of a workflow}
This phase decomposes a workflow graph into smaller data structures that represent sub workflows.
Hence, we created a traverser that explores the workflow graph to gain insight about its complexity and detect its intricate parallel parts.
It obtains information about the workflow inputs, outputs, services, invocations, and associated types.
This information is used to detect the maximum number of smallest sub workflows, each of which consists of a single invocation, or multiple sequential invocations to the same service if a data dependency exists between them.

\subsubsection{Placement analysis}\label{Placement}
\begin{figure}[h]
\centering
 \subfloat[Discovery]{
   \includegraphics[scale=0.8]{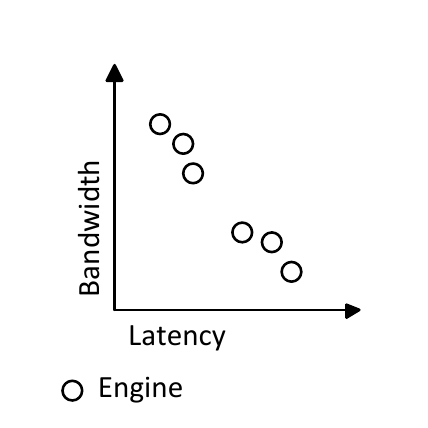}
   \label{discover}
 }
\subfloat[Clustering]{
   \includegraphics[scale=0.8]{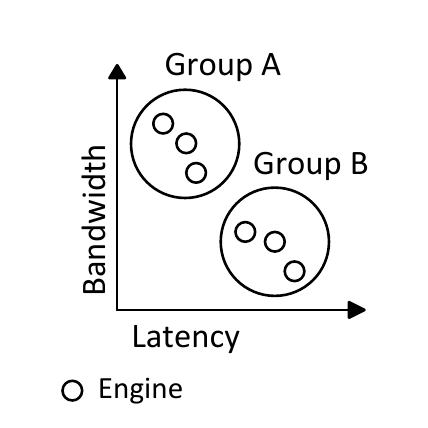}
   \label{clustering}
 }\\
 \subfloat[Elimination]{
   \includegraphics[scale=0.8]{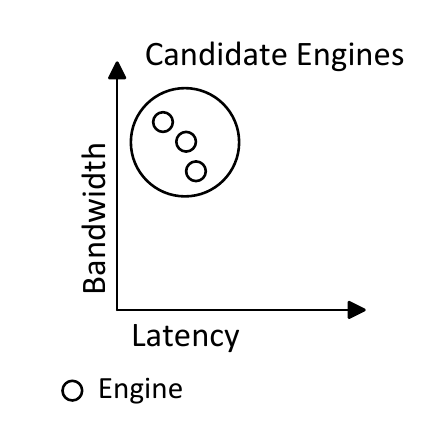}
   \label{elimination}
 }
  \subfloat[Ranking and Selection]{
   \includegraphics[scale=0.8]{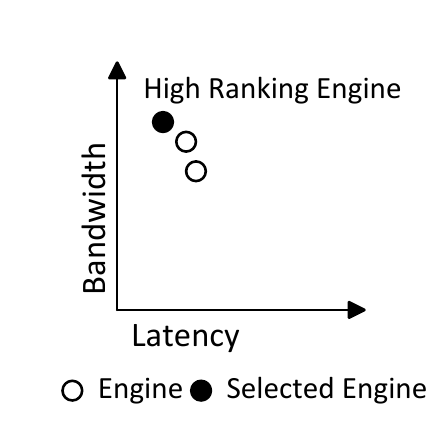}
   \label{ranking}
 }
\caption{Placement analysis.}
\label{workflowplacementfigure}
\end{figure}

Once a workflow has been decomposed by the traverser, placement analysis are performed to determine the most appropriate engines that may execute the sub workflows.
This phase involves the following activities, which are illustrated in figure \ref{workflowplacementfigure}.
\begin{itemize}
\item \textbf{Discovery and clustering of engines:}
This activity identifies a set of available engines that may execute the sub workflows\footnote{This paper does not focus on mechanisms to discover the engines.}.
For each sub workflow, these engines are organised into groups using the k-means clustering algorithm, and according to QoS metrics that represent the network delay, which include the network latency and bandwidth between each engine and the single service endpoint in the sub workflow.
\item \textbf{Elimination of inappropriate engines:}
Upon clustering, groups containing inappropriate engines are eliminated from further analysis.
This is achieved by identifying the engines with metrics that are worse than those of engines in other groups.
\item \textbf{Ranking and selection of engines:}
Each remaining candidate engine is ranked by predicting the transmission time between the engine and the service endpoint using:
\begin{equation} T = L_{e-s} + S_{input}/B_{e-s}
\label{TransmissionTime} \end{equation}
where \begin{math} T \end{math} is the transmission time, \begin{math} L_{e-s} \end{math} and \begin{math} B_{e-s} \end{math} are the latency and bandwidth between the engine and the service respectively, and \begin{math} S_{input} \end{math} is the size of the input that is used to invoke the service.
Consequently, an engine with the shortest transmission time is selected.
\end{itemize}

\subsubsection{Composition of sub workflows}\label{compositionsection}
The sub workflows may be combined together if the same engine is selected to execute them.
This involves introducing directed edges between them wherever a data dependency exists.
Consequently, the composite workflows are encoded using the same language as used to specify the entire workflow.
During the recoding, relevant information such as the workflow inputs, outputs, service invocations, data dependencies and type representations are all captured, and associated with the composite workflows to make each a self contained stand-alone workflow specification.

\subsection{Deployment and Monitoring}

Our approach uses the knowledge about the network condition with a combination of heuristics for initially deploying the workflow.
Each composite workflow specification is dispatched to a designated engine, which compiles and executes it immediately.
This deployment process is transparent and does not require any user intervention.
Upon deployment, real-time distributed monitoring may be used to guide the workflow toward optimal performance.
This is achieved by detecting the network condition periodically and performing further placement analysis.
Our approach uses the application layer capabilities to deliver useful information about the network condition in terms of network latency and bandwidth.
For instance, an engine measures the latency by computing the average round-trip time of a series of HTTP HEAD requests issued to a service.
Similarly, the bandwidth is measured using the request completion time and the response message size.

\section{Workflow Partitioning Example}\label{WorkflowPartitioningExample}

\begin{figure*}[t]
\centering
 \subfloat[]{
   \includegraphics[scale=0.9]{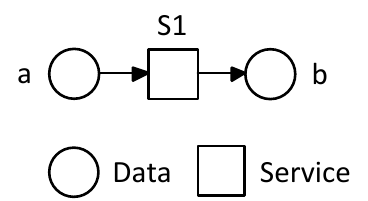}
   \label{subworkflow1}
 }
 \subfloat[]{
   \includegraphics[scale=0.9]{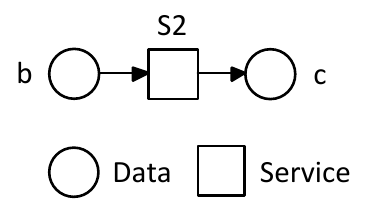}
   \label{subworkflow2}
 }
  \subfloat[]{
   \includegraphics[scale=0.9]{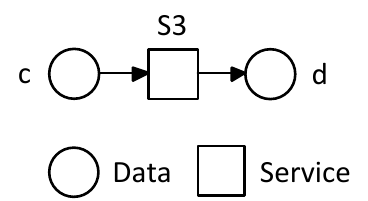}
   \label{subworkflow3}
 }\\
  \subfloat[]{
   \includegraphics[scale=0.9]{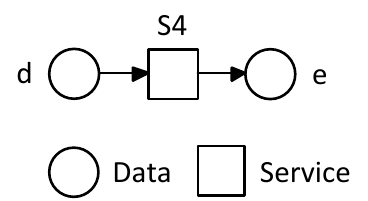}
   \label{subworkflow4}
 }
  \subfloat[]{
   \includegraphics[scale=0.9]{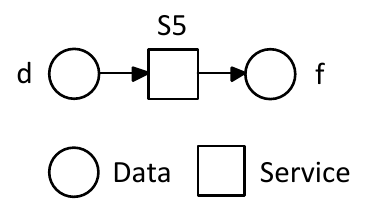}
   \label{subworkflow5}
 }
  \subfloat[]{
   \includegraphics[scale=0.9]{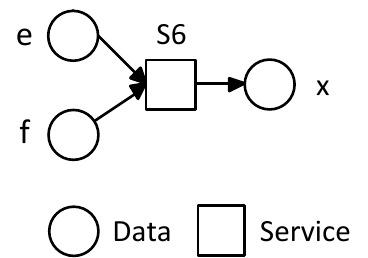}
   \label{subworkflow6}
 }
\caption{Sub workflows obtained from decomposing the workflow shown in figure \ref{workflowfigure}.}
\label{decomposition}
\end{figure*}

This section presents an arbitrary workflow partitioning scenario based on the workflow example in figure  \ref{workflowfigure}, where the workflow is decomposed into smaller sub workflows as shown in figure \ref{decomposition}.
Each sub workflow consists of a single service invocation, which requires one or more inputs and produces a single output.
Upon selecting appropriate engines to execute the sub workflows, they may be combined together to form composite workflows.
During their composition, they are analysed to detect any external dependency between them where an output of a sub workflow is required as an input for another.
Upon detecting an external dependency, it is replaced by a direct service composition between the service endpoint that produces the data and the one that requires it in the composite workflow.
The intermediate data between the services may be represented in the composite workflow as output data when it is required for executing other composite workflows.\\

\begin{figure}[h]
\centerline{\includegraphics[scale=0.9]{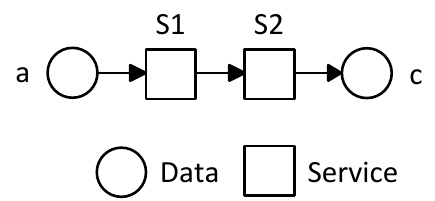}}
\caption{The first composite workflow formed by composing the sub workflows shown in figures \ref{subworkflow1} and \ref{subworkflow2}.}
\label{first}
\end{figure}

Listing \ref{listing2} shows a computer generated specification of the composite workflow shown in figure \ref{first}.
This specification is executed by an engine that is deployed closer to services \begin{ppl}S1\end{ppl} and \begin{ppl}S2\end{ppl} as shown in figure \ref{architecture}.
It shows a universally unique identifier that is specified using the \begin{ppl}\textbf{uid}\end{ppl} keyword in line 2.
This identifier is generated to distinguish the workflow from others with the same name.
The \begin{ppl}\textbf{engine}\end{ppl} keyword declares a remote engine identifier in line 3.
The identifiers relating to the services are all declared through lines 4-9.
The workflow interface is defined through lines 10-13.
The input \begin{ppl}a\end{ppl} is used to invoke \begin{ppl}p1.Op1\end{ppl}, whose output is passed to \begin{ppl}p2.Op2\end{ppl} that produces \begin{ppl}c\end{ppl}.
Finally, the workflow output is forwarded to \begin{ppl}e2\end{ppl} to perform further computation using the \begin{ppl}\textbf{forward}\end{ppl} keyword.

\begin{lstlisting}[keywords={workflow, forward, uid, sid, engine, description, service, port, engine, is, input, output, int, any, transmit, to}, caption={Specification of the first composite workflow shown in figure \ref{first}.}, label={listing2}, frame=single]
01 workflow example
02 uid 618e65607dc47807a51a4aa3211c3298fd8.1
03 engine e2 is http://ec2-54-83-2-120.compute-1.amazonaws.com/services/Engine
04 description d1 is http://ward.host.cs.st-andrews.ac.uk/documents/service1.wsdl
05 description d2 is http://ward.host.cs.st-andrews.ac.uk/documents/service2.wsdl
06 service s1 is d1.Service1
07 service s2 is d2.Service2
08 port p1 is s1.Port1
09 port p2 is s2.Port2
10 input: 
11   int a
12 output:
13   int c
14 a -> p1.Op1
15 p1.Op1 -> p2.Op2
16 p2.Op2 -> c
17 forward c to e2
\end{lstlisting}

\begin{figure}[h]
\centerline{\includegraphics[scale=0.9]{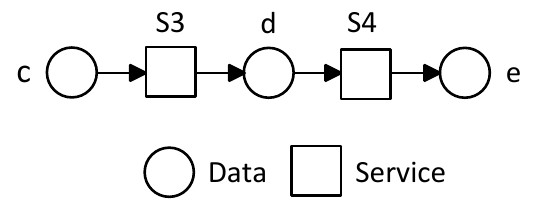}}
\caption{The second composite workflow formed by composing the sub workflows shown in figures \ref{subworkflow3} and \ref{subworkflow4}.}
\label{second}
\end{figure}

Listing \ref{listing3} shows the specification of the second workflow which is shown in figure \ref{second}.
This specification is executed closer to \begin{ppl}s3\end{ppl} and \begin{ppl}s4\end{ppl} by engine \begin{ppl}e2\end{ppl} as shown in figure \ref{architecture}, and upon the availability of the workflow input \begin{ppl}c\end{ppl}.

\begin{lstlisting}[keywords={uid, future, workflow, forward, engine, description, service, port, engine, is, input, output, int, any, transmit, to}, caption={Specification of the second composite workflow shown in figure \ref{second}.}, label={listing3}, frame=single]
01 workflow example
02 uid 618e65607dc47807a51a4aa3211c3298fd8.2
03 engine e3 is http://ec2-54-80-6-125.compute-1.amazonaws.com/services/Engine
04 description d3 is http://ward.host.cs.st-andrews.ac.uk/documents/service3.wsdl
05 description d4 is http://ward.host.cs.st-andrews.ac.uk/documents/service4.wsdl
06 service s3 is d3.Service3
07 service s4 is d4.Service4
08 port p3 is s3.Port3
09 port p4 is s4.Port4
10 input: 
11    int c
12 output:
13    int d, e
14 c -> p3.Op3
15 p3.Op3 -> d
16 d -> p4.Op4
17 p4.Op4 -> e
18 forward d to e3
19 forward e to e3
\end{lstlisting}

\begin{figure}[h]
\centerline{\includegraphics[scale=0.9]{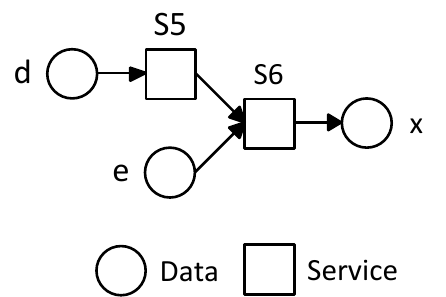}}
\caption{The third composite workflow formed by composing the sub workflows shown in figures \ref{subworkflow5} and \ref{subworkflow6}.}
\label{third}
\end{figure}

Listing \ref{listing4} shows the specification of the workflow in figure \ref{third}, where \begin{ppl}p5.Op5\end{ppl} and \begin{ppl}p6.Op6\end{ppl} are invoked consecutively.
Finally, the workflow output \begin{ppl}x\end{ppl} is forwarded to engine \begin{ppl}e1\end{ppl}, which acts as a data sink for the workflow outputs.
Typically, this engine is the initial engine that partitioned the workflow and deployed it.

\begin{lstlisting}[keywords={workflow, forward, uid, engine, future, description, service, port, engine, is, input, output, int, any, transmit, to}, caption={Specification of the third composite workflow shown in figure \ref{third}.}, label={listing4}, frame=single]
01 workflow example
02 uid 618e65607dc47807a51a4aa3211c3298fd8.3
03 engine e1 is http://ec2-54-80-3-122.compute-1.amazonaws.com/services/Engine
04 description d5 is http://ward.host.cs.st-andrews.ac.uk/documents/service5.wsdl
05 description d6 is http://ward.host.cs.st-andrews.ac.uk/documents/service6.wsdl
06 service s5 is d5.Service5
07 service s6 is d6.Service6
08 port p5 is s5.Port5
09 port p6 is s6.Port6
10 input: 
11    int d, e
12 output:
13    int x
14 d -> p5.Op5
15 p5.Op5 -> p6.Op6.par2
16 e -> p6.Op6.par1
17 p6.Op6 -> x
18 forward x to e1
\end{lstlisting}

\section{Implementation and Evaluation}\label{Evaluation}
The overall implementation is based on Java and wrapped as a web service package that can be deployed on any physical or virtual machine.
It relies upon the Java Runtime Environment (JRE) and Apache Tomcat server.
We have designed a set of experimental workflows to evaluate our approach, each of which is specified based on a particular dataflow pattern.
These workflows consist of a different number of services that are deployed on Amazon EC2, and across several network locations in geographical regions to explore the scalability of our approach.
They are categorised into continental and inter-continental workflows.

\subsection{Configuration}

\subsubsection{Continental workflows}
These workflows consist of services hosted in the same network region such as N. Virginia (us-east-1) over Amazon EC2.
They are orchestrated in the following configurations:

\begin{itemize}
\item \textbf{Centralised orchestration (local):}
The workflow is executed by a workflow engine that is deployed in the same network region where the services are resident as shown in figure \ref{regionallocal}.
\begin{figure}[h]
\centerline{\includegraphics[scale=0.9]{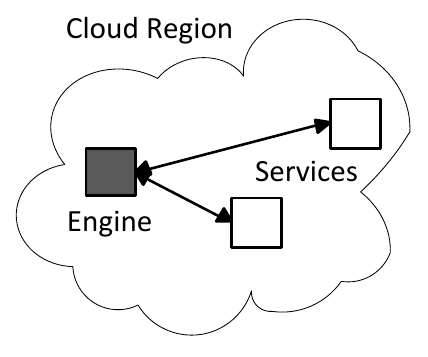}}
\caption{Centralised orchestration (local).}
\label{regionallocal}
\end{figure}
\item \textbf{Centralised orchestration (remote):}
The workflow is executed by a workflow engine that is deployed in a different network region than the one where the services are resident such as N. California (us-west-1).
This configuration is shown in figure \ref{regionalremote}.
\begin{figure}[h]
\centerline{\includegraphics[scale=0.9]{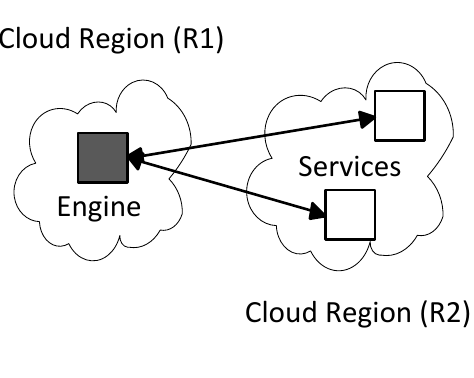}}
\caption{Centralised orchestration (remote).}
\label{regionalremote}
\end{figure}
\item \textbf{Distributed orchestration:}
The workflow is executed using our approach by distributed workflow engines, which are deployed in the same network region where the services are resident as shown in figure \ref{regionaldistributed}.
\begin{figure}[h]
\centerline{\includegraphics[scale=0.9]{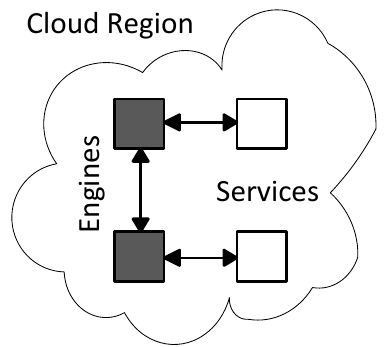}}
\caption{Distributed orchestration of continental services.}
\label{regionaldistributed}
\end{figure}
\end{itemize}

\subsubsection{Configuration of inter-continental workflows}
These workflows consist of services that are distributed across N. Virginia (us-east-1), N. California (us-west-1), Oregon (us-west-2) and Ireland (eu-west-1).
They are orchestrated in the following configurations:
\begin{itemize}
\item \textbf{Centralised orchestration:}
The workflow is executed by a centralised engine that is deployed at an arbitrary network location as shown in figure \ref{continentalcentralised}.
\begin{figure}[h]
\centerline{\includegraphics[scale=0.9]{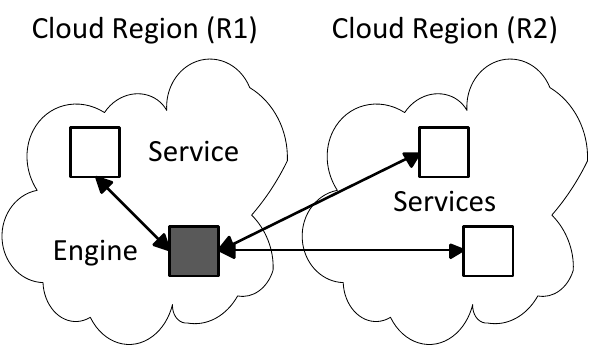}}
\caption{Centralised orchestration.}
\label{continentalcentralised}
\end{figure}
\item \textbf{Distributed orchestration:}
The workflow is executed using our approach by distributed engines that are dispersed over several network regions as shown in figure \ref{continentaldistributed}.

\begin{figure}[h]
\centerline{\includegraphics[scale=0.9]{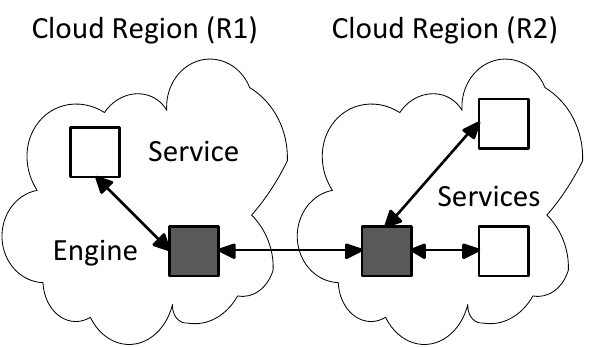}}
\caption{Distributed orchestration.}
\label{continentaldistributed}
\end{figure}
\end{itemize}

\begin{figure*}[t]
\centering
\subfloat[Pipeline Dataflow Pattern (8 Services)]{
   \includegraphics[scale= 0.44]{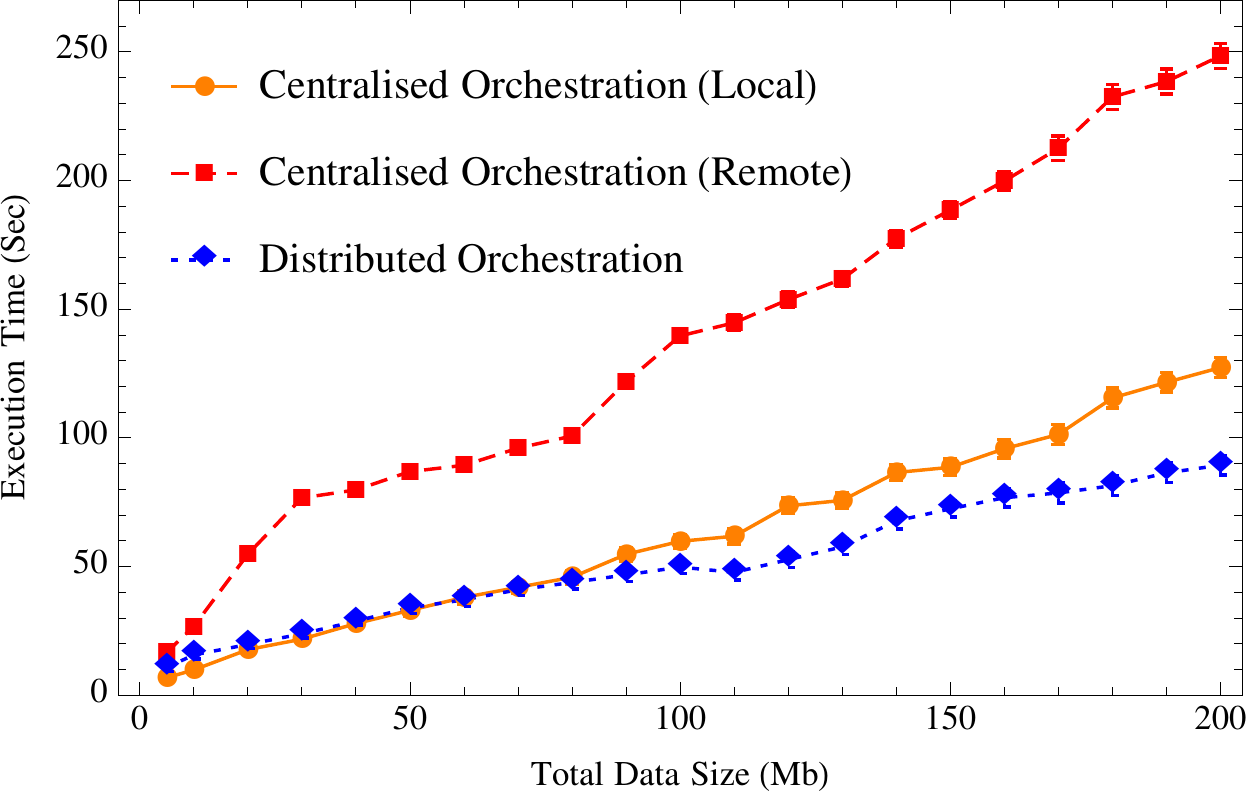}
   \label{regionalsequential8}
 }
 \subfloat[Data Distribution Pattern (8 Services)]{
   \includegraphics[scale= 0.45]{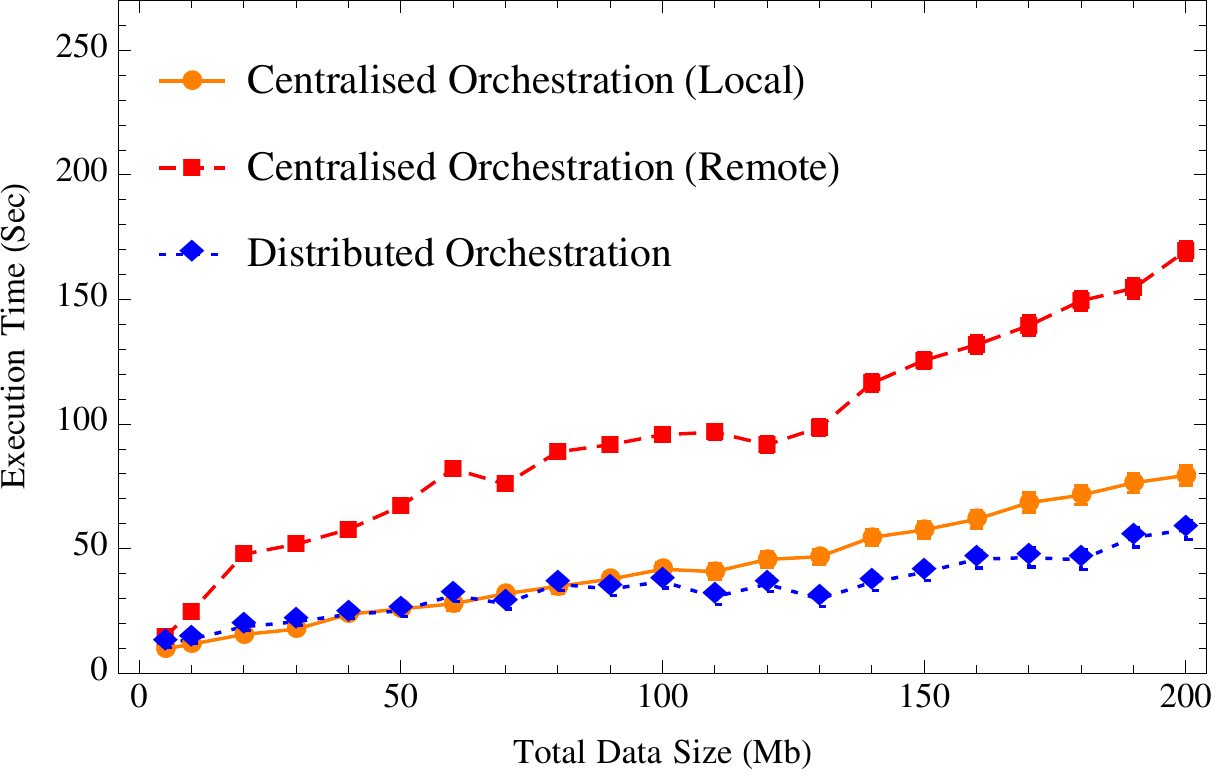}
   \label{regionalfanout8}
 }
 \subfloat[Data Aggregation Pattern (8 Services)]{
   \includegraphics[scale= 0.45]{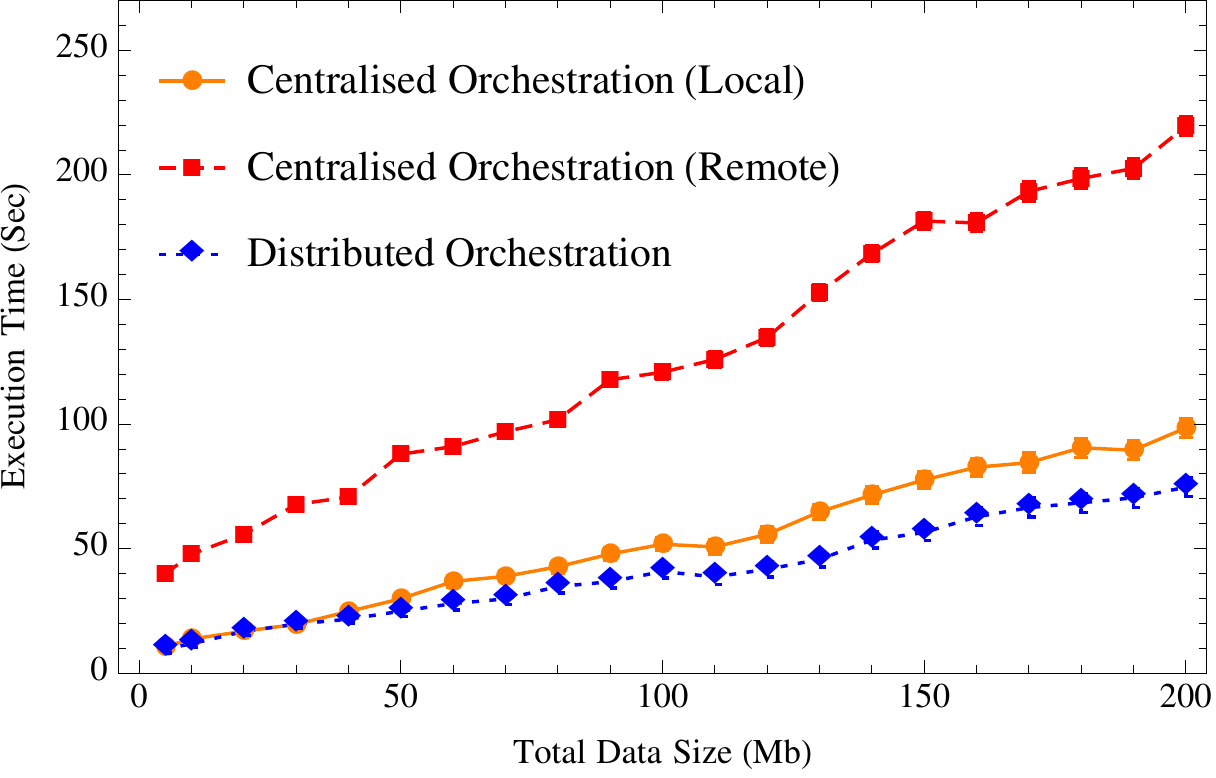}
   \label{regionalfanin8}
 }\\
 \subfloat[Pipeline Dataflow Pattern (16 Services)]{
   \includegraphics[scale= 0.44]{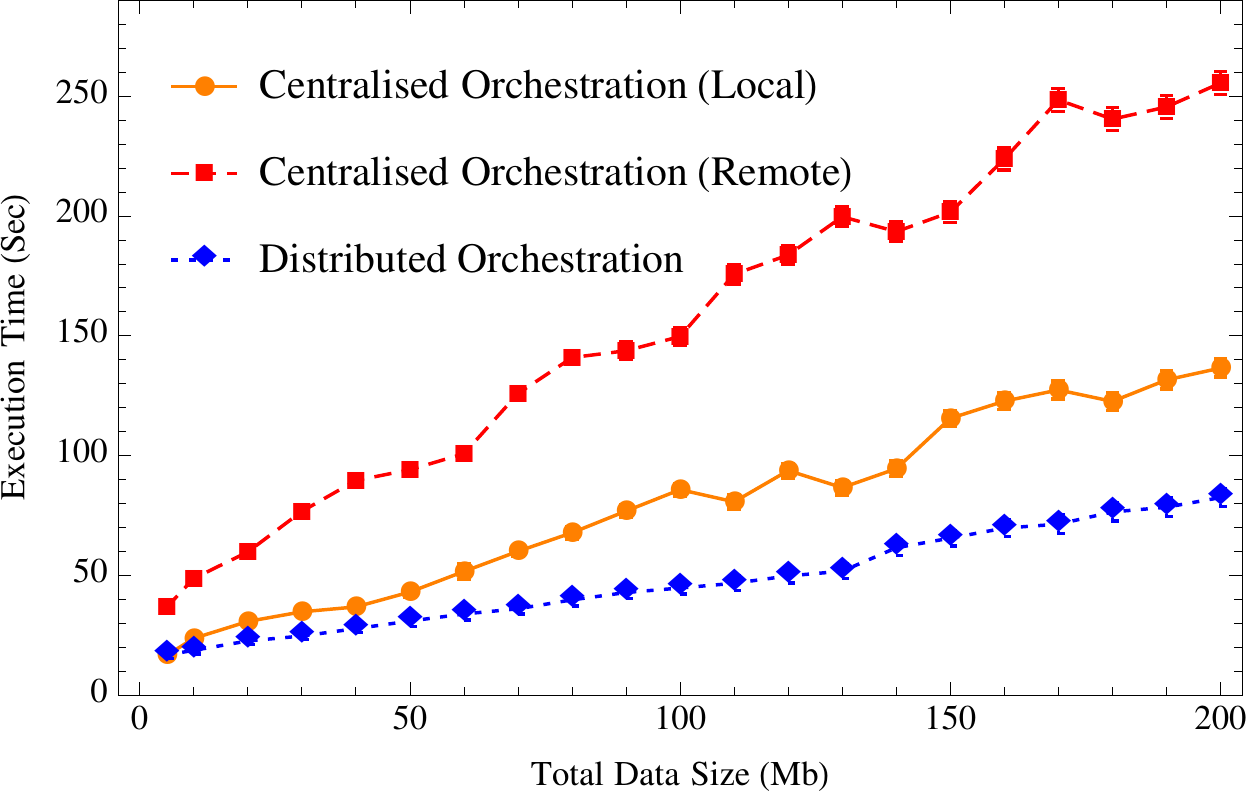}
   \label{regionalsequential16}
 }
 \subfloat[Data Distribution Pattern (16 Services)]{
   \includegraphics[scale= 0.44]{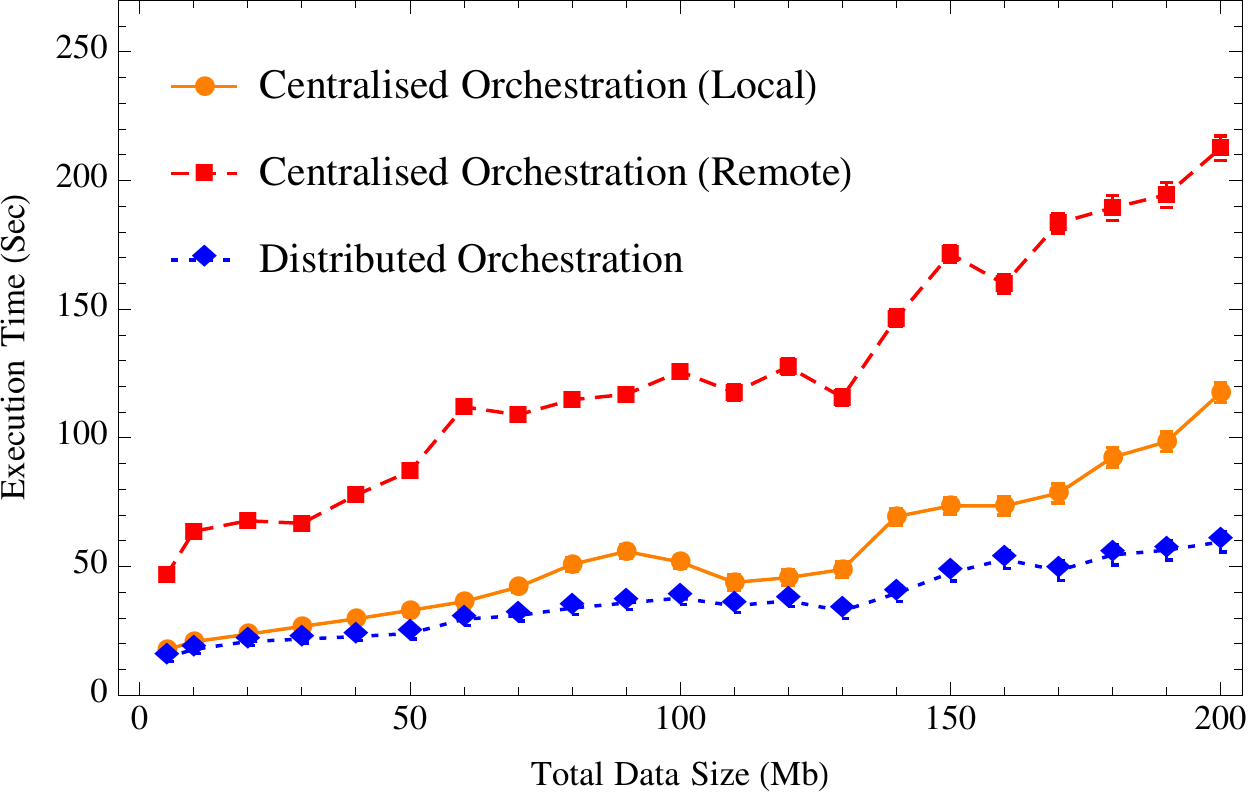}
   \label{regionalfanout16}
 }
 \subfloat[Data Aggregation Pattern (16 Services)]{
   \includegraphics[scale= 0.45]{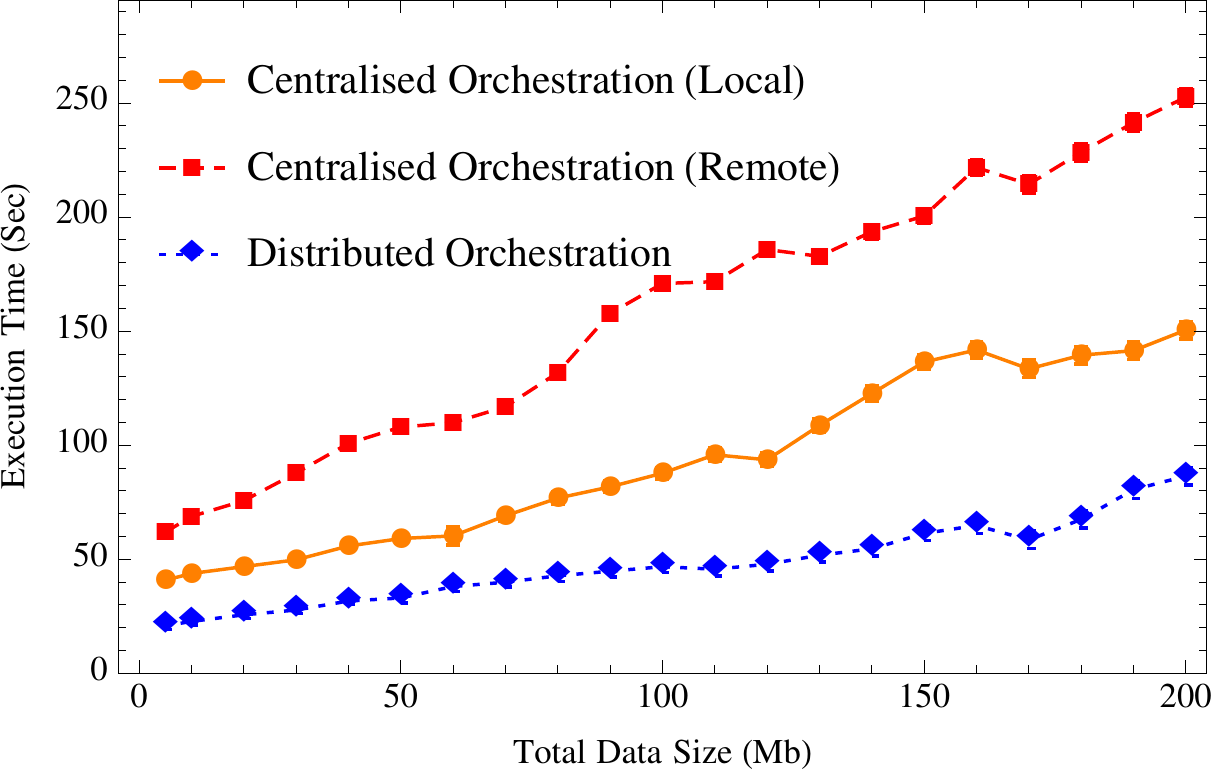}
   \label{regionalfanin16}
 }
\caption{Continental workflow results.}
\label{regionalworkflows}
\end{figure*}

\begin{figure*}[t]
\centering
\subfloat[Pipeline Dataflow Pattern (16 Services)]{
   \includegraphics[scale= 0.44]{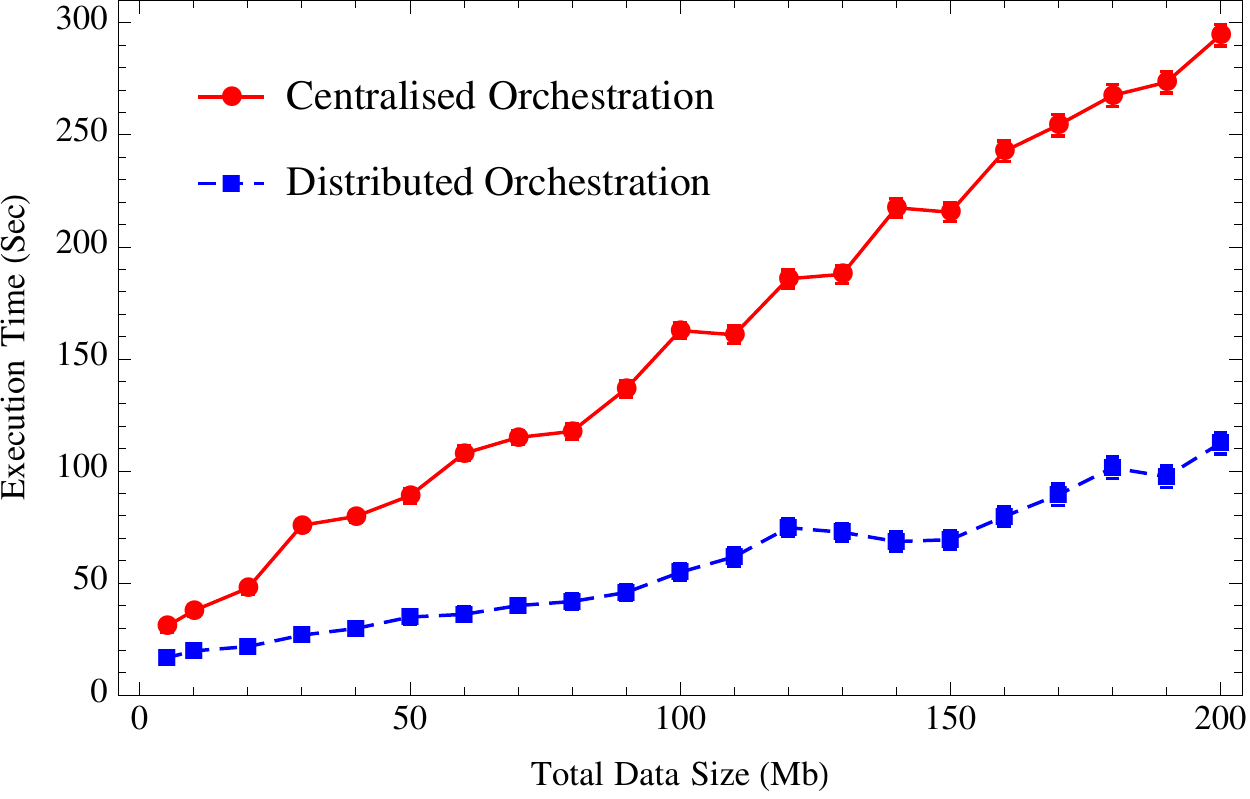}
   \label{continentalsequential8}
 }
 \subfloat[Data Distribution Pattern (16 Services)]{
   \includegraphics[scale= 0.45]{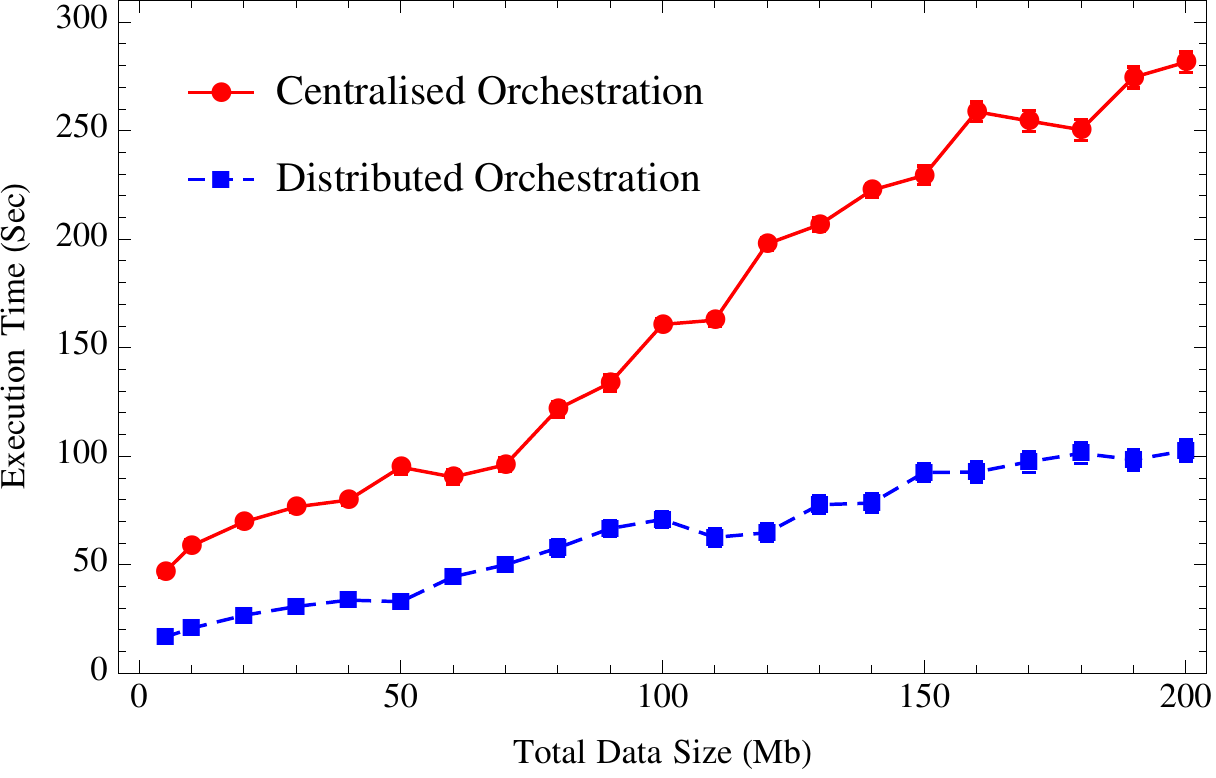}
   \label{continantalsequential16}
 }
\subfloat[Data Aggregation Pattern (16 Services)]{
   \includegraphics[scale= 0.44]{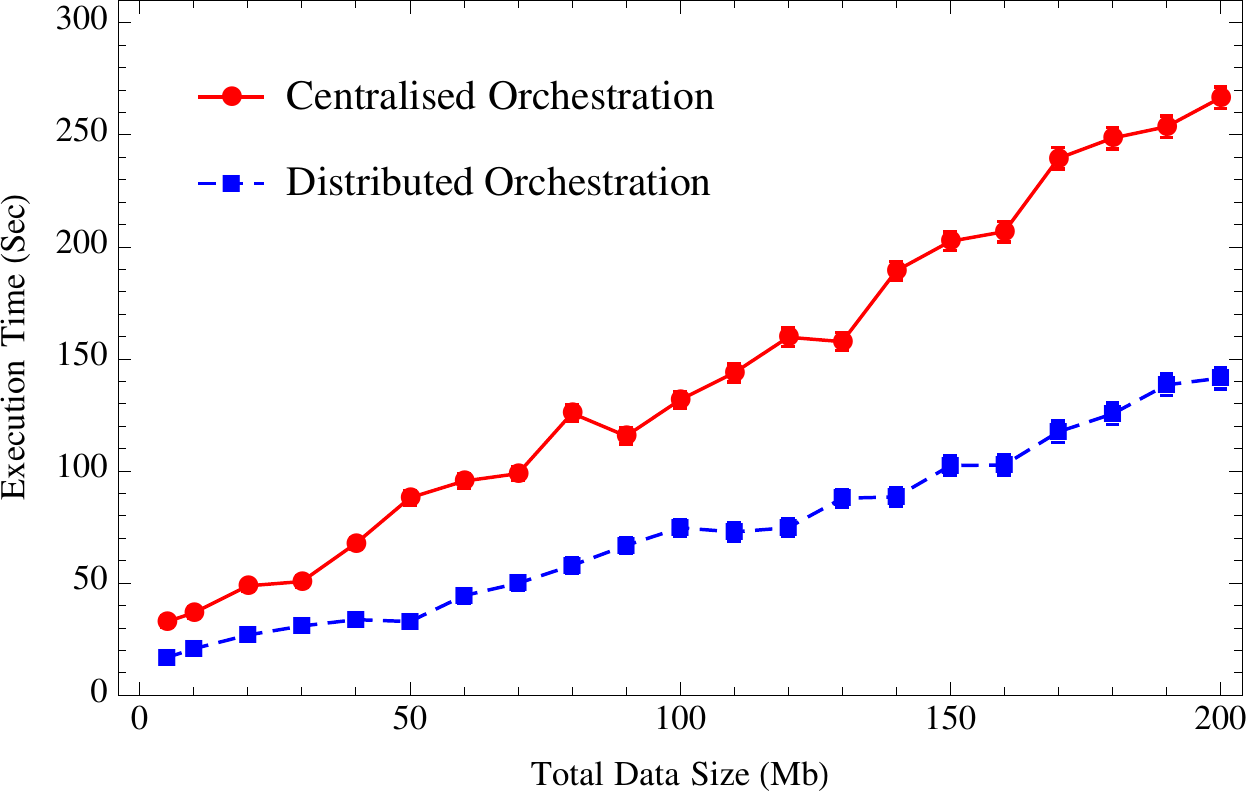}
   \label{continentalfanout8}
 }
\caption{Inter-continental workflow results.}
\label{contworkflows}
\end{figure*}

\subsection{Analysis}

The completion time for each workflow is recorded in seconds, and the size of total communicated data in MB.
Each workflow is executed using 21 inputs to emulate the data increase in each run, and for 20 times (420 runs in total).
The mean speedup rate is computed using:
\begin{equation} S = T_{c} / T_{d}
\label{TransmissionTime} \end{equation} where \begin{math} T_{c} \end{math} and \begin{math} T_{d} \end{math} are the average workflow completion times using centralised and distributed orchestration respectively.\\

\subsubsection{Analysis of continental workflows}
There are a number of observations that can be derived from our experimental results.
Firstly, executing a continental workflow by a centralised engine within the same region provides better performance compared to executing the same workflow by a centralised engine that resides in a remote region.
This is evident in all continental workflows as shown in figure \ref{regionalworkflows}.
Secondly, executing a continental workflow that consists of a small number of services using distributed orchestration may not provide significant performance improvement over local centralised orchestration as shown in figures \ref{regionalsequential8}, \ref{regionalfanout8}, and \ref{regionalfanin8}.
This is because introducing more engines involves the communication of additional intermediate copies of data between them, which may increase the workflow execution time.
Finally, distributed orchestration becomes more useful as the number of services increases according to figures \ref{regionalsequential16}, \ref{regionalfanout16}, and \ref{regionalfanin16}.
Tables \ref{table1}, and \ref{table2} summarise the results where N is the number of services, \begin{math} S_\alpha \end{math} and \begin{math} S_\beta \end{math} are the mean speedup rates for distributed orchestration compared to local and remote centralised orchestration respectively.\\

\begin{table}[h]\normalsize
\caption{Mean speedup rates for continental workflows consisting of 8 services}
\centering
\label{table1}
\begin{center}
\begin{tabular}{| l | l | l | l |}
\hline
Pattern & N & \begin{math} S_\alpha \end{math} & \begin{math} S_\beta \end{math}\\ \hline
Pipeline& 8& 1.13&2.60\\ \hline
Distribution& 8& 1.18&2.69\\ \hline
Aggregation& 8& 1.25&3.23\\ \hline
\end{tabular}
\end{center}
\end{table}

\begin{table}[h]\normalsize
\caption{Mean speedup rates for continental workflows consisting of 16 services}
\centering
\label{table2}
\begin{center}
\begin{tabular}{| l | l | l | l |}
\hline
Pattern & N & \begin{math} S_\alpha \end{math} & \begin{math} S_\beta \end{math}\\ \hline
Pipeline& 16&1.59&3.19\\ \hline
Distribution& 16& 1.43&3.45\\ \hline
Aggregation& 16& 1.93&3.28\\ \hline
\end{tabular}
\end{center}
\end{table}

\subsubsection{Analysis of inter-continental workflows}

Our distributed orchestration approach provides significant performance improvement for all inter-continental workflows as shown in figure \ref{contworkflows}.
Firstly, a centralised engine may take considerable amount of time to execute a workflow due to high latency and low bandwidth between itself and the services.
Secondly, executing a workflow using distributed engines reduces the overall execution time as the data size increases.
This is because several parts of the workflow logic are executed in parallel at the nearest locations to the services, which improves the response times between the engines and the services.
Finally, the time for transferring the intermediate data between the engines may be affected because of the change in the network condition, but it does not degrade the overall workflow performance.
Table \ref{table3} provides the workflow results.

\begin{table}[h]\normalsize
\caption{Mean speedup rates for inter-continental workflows consisting of 16 services}
\label{table3}
\begin{center}
\begin{tabular}{| l | l | l | l |}
\hline
Pattern & N & \begin{math} S \end{math}\\ \hline
Pipeline& 16& 2.69\\ \hline
Distribution& 16&2.54\\ \hline
Aggregation& 16& 1.97\\ \hline
\end{tabular}
\end{center}
\end{table}

\begin{figure}[!hb]
\centerline{\includegraphics[scale=0.55]{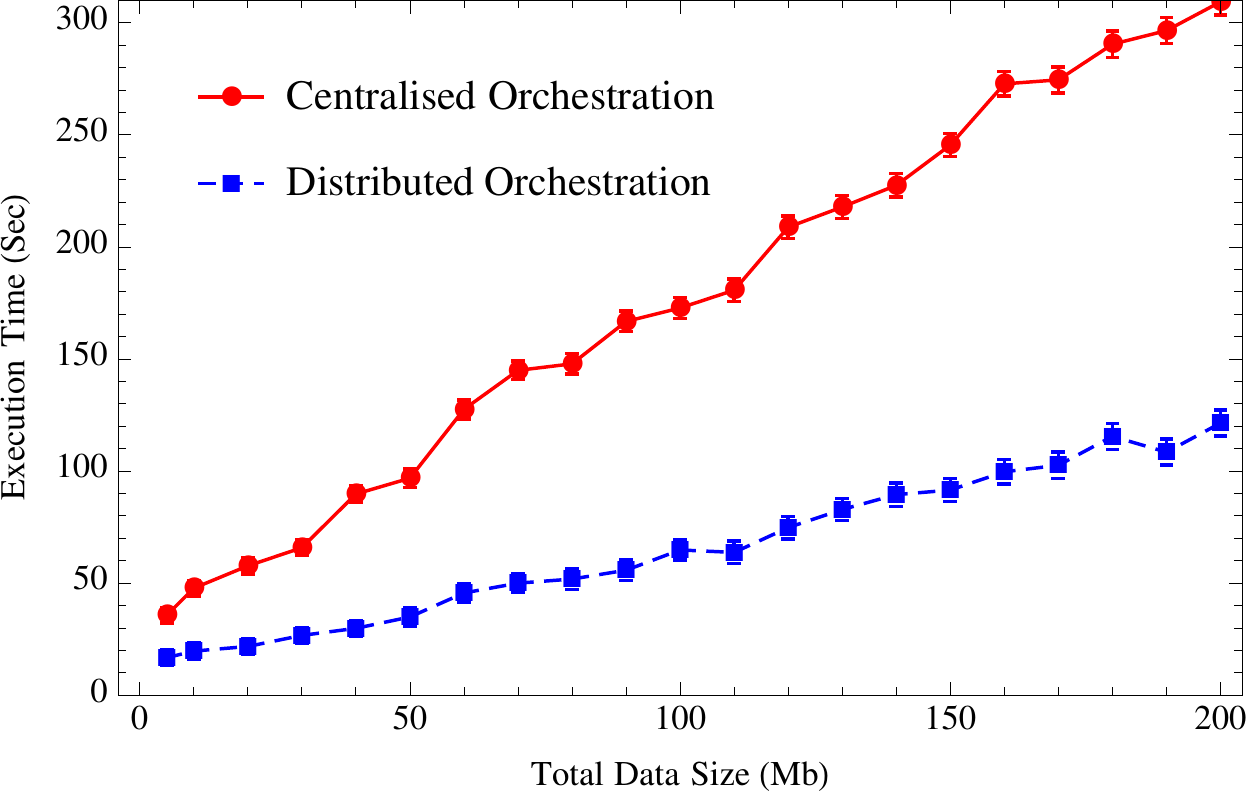}}
\caption{End-to-end inter-continental workflow.}
\label{endtoendworkflow}
\end{figure}
\subsubsection{Analysis of an inter-continental end-to-end workflow}
Although this paper has focused primarily on evaluating our approach based on common dataflow patterns, it is essential to demonstrate its efficacy based on an end-to-end workflow application that combines all these patterns together.
Hence, we created a workflow that consists of 16 services which are distributed across multiple regions.
Figure \ref{endtoendworkflow} shows the overall results where the mean speedup is 2.68.
The final outputs of all inter-continental workflows are obtained from the services, and stored on machines that host the engines which obtained the outputs.

\section{Related Works}\label{RelatedWorks}

\subsection{Workflow Management Systems}
Pegasus \cite{Pegasus} is a notable workflow management system that relies on a centralised engine for scheduling distributed workflow tasks.
Our approach is based on a distributed design model that permits the workflow partitions to be executed without prior scheduling.
Each sub workflow is executed automatically as soon as the data that is required for its execution is available from other sources.
Condor \cite{Condor} leverages resource machines for executing distributed tasks using batch processing.
These tasks are specified manually by the user, and pre-knowledge about the machines and the condition of the network is required.
Our architecture handles the partitioning and mapping of workflows onto machines automatically by collecting information about the network condition, and performing placement analysis.
Triana \cite{Triana} permits a workflow to be distributed across machines, and supports control flows by associating coordination logic with workflow scripts.
Control flows are unnecessary in our approach as it is relies on a dataflow language that avoids loops and control structures in the workflow.
Kepler \cite{Kepler} is based on actor-oriented modelling that permits a workflow to be composed using actors which communicate through well-defined interfaces.
However, it does not support decentralised execution of workflows.

\subsection{Dataflow Optimisation Architectures}
The Circulate approach \cite{Circulate}, \cite{Circulate2} supports data distribution in the workflow using proxy components, which may be deployed closer to the services.
These proxies exploit connectivity to the services, and route the data in the workflow to locations where they are required.
However, this architecture relies on a centralised flow mechanism to facilitate the collaboration between proxies, and there seems to be no automated mechanism for partitioning the workflow.
The Flow-based Infrastructure for Composing Autonomous Services (FICAS) \cite{Ficas} supports service composition by altering the service interfaces to enable peer-to-peer collaboration.
Our approach does not require modifying the implementation of services.
Data processing techniques built on MapReduce \cite{MapReduce} may be suitable for a wide range of problems, but are inadequate for executing workflows.
For instance, a workflow can be composed using a series of MapReduce jobs \cite{MapReduce2}, but this requires passing the entire state and data from one job to the next which degrades performance.
Dryad \cite{Dryad} supports executing distributed processing tasks, but it does not provide any mechanism to rearrange the workflow structure for optimisation purposes.
Furthermore, the distributed workflow parts must be specified manually.
Our approach automatically generates the distributed sub workflow specifications.

\subsection{Workflow Scheduling Approaches}
There are many scheduling heuristics that attempt to solve the workflow mapping problem such as HEFT \cite{HEFT}, Min-Min \cite{MinMin}, MaxMin and MCT \cite{MaxMinMCT}, but these works are directed at grid-based workflow applications.
Several other heuristic methods were proposed and compared in \cite{DAGSchedulingHeuristics}.
Partitioning is proposed for provisioning resources into execution sites in \cite{NonDAG} and \cite{NonDAG2}, but not for decomposing the actual dataflow graph.
In \cite{RunTimeOptimisation} and \cite{Askalon}, a strategy is discussed where the workflow tasks are mapped onto grid sites.
This is achieved by assigning weights to the vertices and edges in the workflow graph by predicting the execution time for each task, and the time for transferring data between the resources.
Each task is then mapped onto a resource that provides the earliest expected time to complete its execution.
However, the time for executing a service operation cannot be predicted efficiently in a service-oriented environment as it depends on the application logic, and the underlying protocols and infrastructure.

\section{Conclusion}\label{Conclusion}

Centralised service orchestration presents significant scalability problems as the number of services and the size of data involved in the workflow increases.
These problems include the unnecessary consumption of the network bandwidth, high latency in transmitting data between the services, and performance bottlenecks.
This paper has presented and evaluated a novel workflow partitioning approach that decomposes a workflow into smaller sub workflows, which may then be transmitted to appropriate locations at which their execution takes place.
These locations are carefully determined using a heuristic technique that relies on the knowledge of the network condition.
This permits the workflow logic to be executed within short geographical distance to the services, which improves the overall workflow performance.
Future work will focus on real-time distributed monitoring, and re-deployment of executing sub workflows to adapt to dynamic changes in the execution environment.

\nocite{*}
\bibliographystyle{ieeetr}
\bibliography{lib}

\end{document}